\documentclass{article}
\usepackage[utf8]{inputenc} % allow utf-8 input
\usepackage[T1]{fontenc}    % use 8-bit T1 fonts
\usepackage[square,sort,comma,numbers]{natbib}
\usepackage{amsmath,amssymb,amsfonts}
\usepackage[noend]{algorithmic}
\usepackage{graphicx}
\usepackage{bm}
\usepackage{textcomp}
\usepackage{xcolor}
\usepackage{multirow}
\usepackage{mathrsfs}
\usepackage{mathtools}
\usepackage{adjustbox}
\usepackage{tabularx,booktabs}
\usepackage{mathtools}
\usepackage{arydshln}
\usepackage{algorithm}
\usepackage{xparse}
\usepackage{caption} 
\usepackage{fullpage}
\usepackage{pifont}

\def\E{\mathbb{E}}

\def\xbm{{\bm{x}}}
\def\zbm{{\bm{z}}}
\def\ybm{{\bm{y}}}
\def\zbm{{\bm{z}}}

\def\thetabm{{\bm{\theta }}}

\def\Ibf{{\mathbf{I}}}

\def\Ncal{{\mathcal{N}}}

\newcommand{\norm}[1]{\left\lVert#1\right\rVert}

\begin{document}

%\title{Generative Diffusion Model Augmented Gradient Echo Plural Contrast Imaging\\}
\title{DiffGEPCI: 3D MRI Synthesis from mGRE Signals \\using 2.5D Diffusion Model}
%\title{DeepGEPCI: Volumetric Magnetic Resonance Image Synthesis from mGRE Signals}

\author{Yuyang Hu,  Satya~V.~V.~N.~Kothapalli, Weijie~Gan, \\Alexander~L.~Sukstanskii, Gregory F. Wu,  Manu Goyal,\\ Dmitriy~A. Yablonskiy, Ulugbek~S.~Kamilov \\
\small{Washington University in St.Louis}\\
%\texttt{Department of Electrical and Systems Engineering, Washington University in St.~Louis}, \texttt{Mallinckrodt Institute of Radiology, Washington University in St.~Louis}, \texttt{Department of Computer Science and Engineering, Washington University in St.~Louis}, \texttt{Department of Neurology, Washington University in St.~Louis School of Medicine}, \texttt{Department of Pathology and Immunology, Washington University in St.~Louis School of Medicine}   \\
}
\maketitle

\begin{abstract}
We introduce a new framework called DiffGEPCI for cross-modality generation in magnetic resonance imaging (MRI) using a 2.5D conditional diffusion model. DiffGEPCI can synthesize high-quality Fluid Attenuated Inversion Recovery (FLAIR) and Magnetization Prepared-Rapid Gradient Echo (MPRAGE) images, without acquiring corresponding measurements, by leveraging multi-Gradient-Recalled Echo (mGRE) MRI signals as conditional inputs. DiffGEPCI operates in a two-step fashion: it initially estimates a 3D volume slice-by-slice using the axial plane and subsequently applies a refinement algorithm (referred to as 2.5D) to enhance the quality of the coronal and sagittal planes. Experimental validation on real mGRE data shows that DiffGEPCI achieves excellent performance, surpassing generative adversarial networks (GANs) and traditional diffusion models.
\end{abstract}

\section{Introduction}
Magnetic Resonance Imaging (MRI) is a versatile medical imaging technique that can produce multiple image contrasts that provide complementary information about a tissue. There is a growing interest in MRI to synthesize multiple naturally co-registered images corresponding to different MRI contrasts from  a single scan~\cite{panda2017magnetic}. Gradient Echo Plural Contrast Imaging (GEPCI)~\cite{luo2012gradient} has been previously proposed for generating FLAIR~\cite{bakshi2001fluid}, MPRAGE~\cite{mugler1991rapid}, and other commonly used contrasts images from a single Multi-Gradient-Recalled Echo (mGRE) sequence data~\cite{elster1993gradient}. However, due to its purely physics-driven nature, original GEPCI algorithms did not provide perceptually high-quality images similar to those used in clinic. We seek to address this issue by proposing a new data-driven method for generating FLAIR and MPRAGE images with high-perceptual quality from mGRE.

Generative models are currently popular for image-to-image translation~\cite{pang2021image}. Conditional generative adversarial networks (cGAN) have been shown to create realistic images~\cite{isola2017image,dar2019image,armanious2020medgan}. The training of cGANs relies on adversarial learning, where a discriminator captures target distribution information and guides a mapping from source to target. Conditional diffusion models (CDMs) have recently surpassed cGANs in terms of ability to generate high-quality and realistic images~\cite{dhariwal2021diffusion, ho2020denoising, saharia2022palette, kazerouni2023diffusion, wang2023zero}. CDMs start with noise and progressively refine output through iterative denoising while also integrating conditional information. This advancement has enabled CDMs to achieve state-of-the-art results in multi-contrast MR synthesis~\cite{ozbey2023unsupervised}. In the context of 3D multi-contrast MRI synthesis, current diffusion models are limited to generating images slice-by-slice. This approach can potentially compromise the preservation of fine details and may introduce artifacts along the other two dimensions. Directly implementing a 3D diffusion model to address this challenge may not be feasible for two primary reasons: (a) it necessitates a significantly larger dataset (comprising thousands of volumes as opposed to thousands of slices) and (b) requires higher computational cost.

We introduce the Denoising Diffusion Model for Gradient Echo Plural Contrast Imaging (DiffGEPCI) method that incorporates a 2.5D processing into diffusion-model-based MRI synthesis. DiffGEPCI uses a conditional diffusion model trained with mGRE signals as conditional input, sourced from three distinct planes: Axial, Coronal, and Sagittal. During inference, DiffGEPCI generates a primary translated volume from the Axial plane and incorporates a specialized 2.5D refinement module to improve performance across all three planes. DiffGEPCI consistently yields state-of-the-art results for both  mGRE $\rightarrow$ FLAIR and mGRE $\rightarrow$ MPRAGE synthesis.

\begin{figure*}[ht]
	\centering
	\includegraphics[width=.98\textwidth]{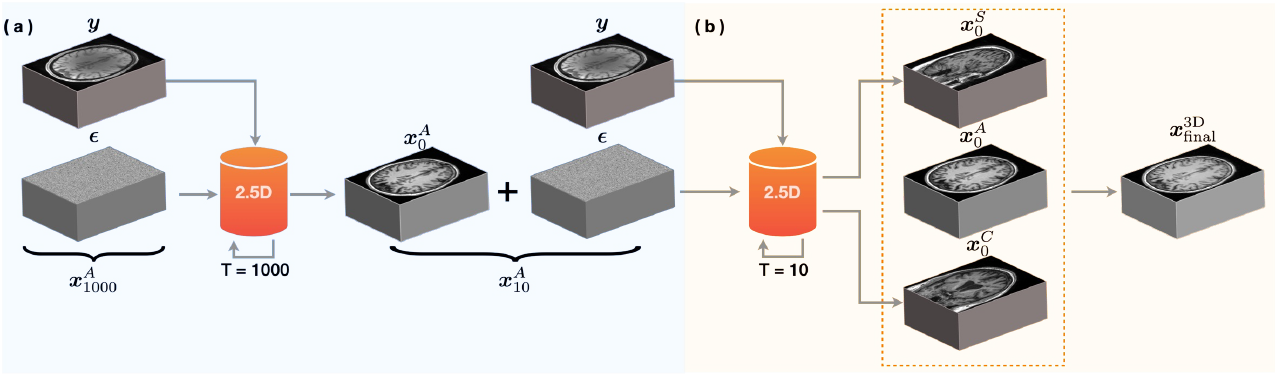}
	\caption{DiffGEPCI relies on a two-step process for enhancing the quality of MRI synthesis. In the first step (a), a initial target modality volume is generated from the axial plane. In the second step (b), this volume is split into two additional plane volumes, and a 10-step inverse diffusion process is applied to improve their quality. Finally, the three volumes are fused by averaging their pixel values, leading to enhanced performance in both coronal and sagittal planes. }
	\label{fig:method}
\end{figure*}

\begin{figure*}[ht]
	\centering
	\includegraphics[width=.98\textwidth]{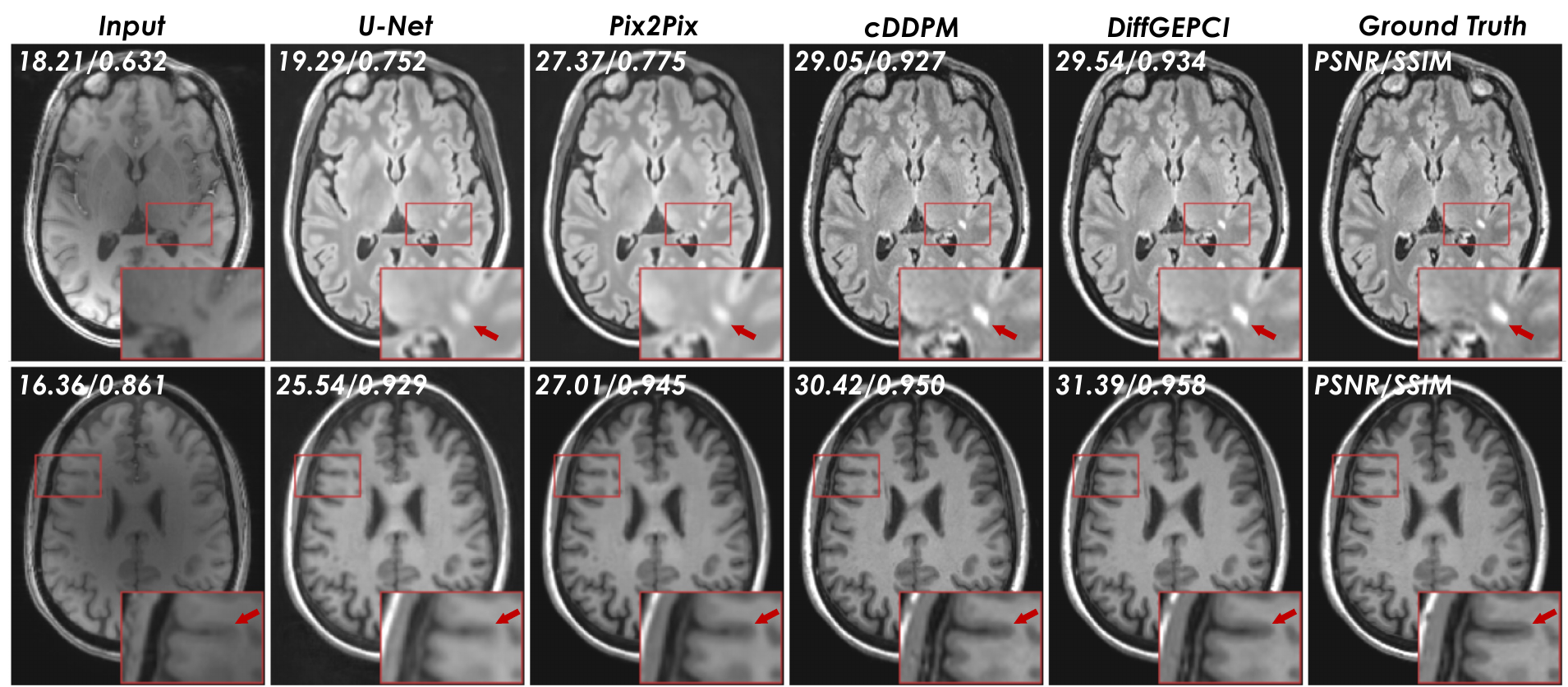}
	\caption{Performance Comparison of several baselines and our proposed method in Aixal planes. The top row illustrates the visual performance mGRE $\rightarrow$ FLAIR task, while the bottom row presents mGRE $\rightarrow$ MPRAGE. Our method estimates intricate structures while avoiding troublesome artifacts, leading to the best PSNR and SSIM performance.}
	\label{fig:results}
\end{figure*}

\section{Proposed Method}

\subsection{Conditional Denoising Diffusion Probabilistic Model}
Conditional Denoising Diffusion Probabilistic Model (cDDPM) is designed to learn a parameterized Markov chain to convert Gaussian distribution to conditional data distribution $p(\xbm, \ybm)$, see~\cite{ho2020denoising,armanious2020medgan} for review of DDPM.  In this paper,  $\xbm$ refers to our target modalities and the condition $\ybm$ refers to the 10  mGRE signals (see~\cite{luo2012gradient} for information on mGRE). The forward diffusion process starts from clean sample $\xbm_0$ and then gradually adds Gaussian noise to $\xbm_0$ according to the following transition probability:
\begin{equation}
    q(\xbm_t|\xbm_{t-1}) := \Ncal{(\xbm_t; \sqrt{1-\beta_t}\xbm_{t-1},\beta_t\Ibf}),
\end{equation}
where $t\in[0, T]$, $\Ncal(\cdot)$ denotes the Gaussian probability density function (pdf), $\beta_{t}$ refers to the noise schedule of the process. By parameter changes $\alpha_t := 1-\beta_{t}$ and $\bar{\alpha}_t := \prod^{t}_{s=1}\alpha_s$, we can rewrite $\xbm_t$ as a linear combination of noise $\bm\epsilon$ and $\xbm_0$
\begin{equation}
    \xbm_t = \sqrt{\bar{\alpha}_t}\xbm_0 + \sqrt{1-\bar{\alpha}_t}\bm\epsilon,
\end{equation}
where $\epsilon \sim \Ncal(0, \Ibf)$. This allows a closed-form expression
for the marginal distribution for sampling $\xbm_t$ given $\xbm_0$
\begin{equation}
    q(\xbm_t|\xbm_{0}) := \Ncal(\xbm_t; \sqrt{\bar{\alpha}_t}\xbm_0, (1-\bar{\alpha}_t)\Ibf).
\end{equation}
The goal of the reverse process is to generate a clean image $\xbm_0$ given a noise vector $\xbm_T$ and the condition $\ybm$. This is achieved by training a convolutional neural network (CNN) to reverse the Markov Chain from $\xbm_T$ to $\xbm_0$. The learning is formulated as the estimation of a parameterized Gaussian process
\begin{equation}
    p(\xbm_{t-1}|\xbm_{t}) := \Ncal(\xbm_{t-1}; \bm{\mu}_\thetabm(\xbm_t, t, \ybm), \sigma^2_t\Ibf),
\end{equation}
where the mean $\bm{\mu}_\thetabm(\xbm_t, t)$ is only uncertain variable we need to estimate and  $\sigma^2_t = \frac{1-\bar{a}_t}{1-\bar{a}_{t-1}}\beta_t$. cDDPM uses a CNN trained to estimate $\bm\epsilon_\thetabm$ to estimate added noise $\bm{\epsilon}$ to reparametrize sampling from the conditional distribution into
\begin{equation}
    \xbm_{t-1} = \frac{1}{\sqrt{a_t}}(\xbm_t - \frac{\beta_t}{\sqrt{1-\bar{a}_t}}\bm\epsilon_\thetabm(\xbm_t,t, \ybm)) + \sigma_t\zbm,
\end{equation} 
 where $\zbm \sim  \Ncal(0, \Ibf)$. The CNN $\bm{\epsilon}_\thetabm$ used in our implementation is learned by minimizing the following expected loss
\begin{equation}
\label{Eq:training}
\mathcal{L}(\thetabm) = \E_{\bm\epsilon, \xbm_t, t} [ \norm{\bm\epsilon_\thetabm(\xbm_t, t, \ybm) - \bm\epsilon}_2^2].
\end{equation}
\subsection{DiffGEPCI for High-Quality Volumetric Synthesis}

We now introduce DiffGEPCI, a new framework specifically designed for translating 3D mGRE signals into the target modalities by using 2.5D conditional diffusion models.

Our data preprocessing standardizes each target and condition volume by padding them to a uniform size across all planes, ensuring that the volume dimension is $N \times N \times N$. We use notations $\xbm^{A_i}$, $\xbm^{C_j}$, and $\xbm^{S_k}$ to denote the 2D slices from the the Axial, Coronal, and Sagittal planes, indexed using $i$, $j$, and $k$. Furthermore, we represent the 3D volumes derived from the Axial, Coronal, and Sagittal planes as  $\xbm^{A}$, $\xbm^{C}$, and $\xbm^{S}$, respectively.
%These slices all originate from the same 3D volume, denoted as $\xbm^{3D}$, which can be expressed as $\xbm^{3D} = \{\xbm^{A_i}\}^N_{i=1} = \{\xbm^{C_j}\}^N_{j=1} = \{\xbm^{S_k}\}^N_{k=1}$. 

DiffGEPCI is trained with three distinct target planes using the same model with the loss function defined in \eqref{Eq:training}, where $\xbm_0 \sim q(\xbm_0)$ and $q$ is the probability distribution of target modalities from three planes. DiffGEPCI uses the same cDDPM model as implemented in~\cite{dhariwal2021diffusion}. This approach enables the use of a single model $\bm\epsilon_\thetabm$ to estimate all planes in the target modality.

As depicted in Figure 1(a), during the inference phase, we initiate the prediction process by estimating each slice of the target modality along the Axial plane in a 1000-step iterative process, as defined in Algorithm~\ref{Alg:inital}. 
Once we predict all the axial slices of the 3D volume, we obtain the initial target volume, denoted as $\xbm_{\text{inital}}^{\text{3D}} =\xbm_0^{A}= (\xbm_0^{A_i})^N_{i=1}$.
Although $\xbm_{\text{inital}}^{\text{3D}}$ provides reasonably good performance, it still exhibits artifacts in the Coronal and Sagittal planes, as shown in Fig.~\ref{fig:z_results}. 

To address inconsistencies along Coronal and Sagital planes, we introduce a subsequent refinement module illustrated in Fig.~\ref{fig:method}(b). First, we split the $\xbm_{\text{inital}}^{\text{3D}}$ into coronal and sagittal planes, creating $\xbm^{C} = (\xbm^{C_j})^N_{j=1}$ and $\xbm^{S} = (\xbm^{S_k})^N_{k=1}$. Next, we apply 10 forward diffusion steps to introduce a minor amount of noise to each slice, resulting in  $\xbm_{10}^{C} = (\xbm_{10}^{C_j})^N_{j=1}$ and $\xbm_{10}^{S} = (\xbm_{10}^{S_k})^N_{i=1}$. Following this, as shown in Algorithm~\ref{Alg:Refinement}, the same model $\bm{\epsilon}_\thetabm$ is used to execute 10 steps of an inverse diffusion process for each slice, yielding $\xbm_{0}^{C}$ and $\xbm_{0}^{S}$. Finally, the three volumes are fused to obtain the final output $\xbm_{\text{final}}^{\text{3D}}$.

\section{Experimental Validation}
\subsection{Dataset}
All human studies were approved by Washington University IRB. We collected high-resolution 3D brain MRI data from 29 volunteers using a Siemens 3T Trio MRI scanner and a 32-channel phased-array head coil, obtaining mGRE, FLAIR, and MPRAGE images at ($1\times1\times1$mm) resolution for each subject to create paired datasets. mGRE data were acquired using a three dimensional (3D) multi-gradient-echo sequence with FOV 256 mm$\times$192 mm, repetition time TR = 50 ms, flip angle 35°, 10 gradient echoes with first gradient echo time TE1 = 4 ms, echo spacing  = 4 ms. Additional phase stabilization echo (the navigator) is collected for each line in k-space to correct for image artefacts due to the physiological fluctuations~\cite{wen2015role}. The Generalized Autocalibrating Partially Parallel Acquisitions (GRAPPA)~\cite{griswold2002generalized} algorithm was used to reduce imaging time to 10 minutes. The 29 subjects were split into 23, 2, and 4 for training, validation, and testing, respectively. We extracted 85 slices from each subject's brain volume, containing the most relevant brain regions. We co-registered the corresponding FLAIR and MPRAGE data to the mGRE sequence using a 3D Affine transformation from the Dipy package~\cite{garyfallidis2014dipy}.

\begin{algorithm}[t]
\caption{: Initial synthesis}
\label{Alg:inital}
\begin{algorithmic}[1]
\STATE \textbf{input:} $\bm\epsilon_\thetabm$: pre-trained model, $\ybm$: mGRE signal

\FOR{$i = N, N-1,\cdots, 0$}
\STATE Sample $\xbm_{1000}^{A_i} \sim  \Ncal(0, \Ibf)$
\FOR{$t = 1000,999,\cdots, 0$}
\STATE $\zbm \sim  \Ncal(0, \Ibf)$
\STATE $\xbm_{t-1}^{A_i} = \frac{1}{\sqrt{a_t}}(\xbm_t^{A_i} - \frac{\beta_t}{\sqrt{1-\bar{a}_t}}\bm{\epsilon}_\thetabm(\xbm_t^{A_i},t, \ybm^{A_i})) + \sigma_t\zbm$.
\ENDFOR
\ENDFOR
\end{algorithmic}
\end{algorithm} 

\begin{algorithm}[t]
\caption{: 2.5D Refinement Module}
\label{Alg:Refinement}
\begin{algorithmic}[1]
\STATE \textbf{input:} $\bm\epsilon_\thetabm$: pre-trained model; $\ybm$: mGRE signal; $\xbm_{0}^{A}$, $\xbm_{10}^{C}$ and  $\xbm_{10}^{S}$: initialization volume,

\FOR{$j = N, N-1,\cdots, 0$}
\FOR{$t = 10,9,\cdots, 0$}
\STATE $\zbm \sim  \Ncal(0, \Ibf)$
\STATE $\xbm_{t-1}^{C_j} = \frac{1}{\sqrt{a_t}}(\xbm_t^{C_j} - \frac{\beta_t}{\sqrt{1-\bar{a}_t}}\bm{\epsilon}_\thetabm(\xbm_t^{C_j},t, \ybm^{C_j})) + \sigma_t\zbm$.
\ENDFOR
\ENDFOR
\FOR{$k = N, N-1,\cdots, 0$}
\FOR{$t = 10,9,\cdots, 0$}
\STATE $\zbm \sim  \Ncal(0, \Ibf)$
\STATE $\xbm_{t-1}^{S_k} = \frac{1}{\sqrt{a_t}}(\xbm_t^{S_k} - \frac{\beta_t}{\sqrt{1-\bar{a}_t}}\bm{\epsilon}_\thetabm(\xbm_t^{S_k},t, \ybm^{S_k})) + \sigma_t\zbm$.
\ENDFOR
\ENDFOR
\RETURN $\xbm_{\text{final}}^{\text{3D}} = \frac{1}{3}(\xbm_{0}^{A} + \xbm_{0}^{C} + \xbm_{0}^{S}).$
\end{algorithmic}
\end{algorithm} 

%\begin{figure*}[ht]
%	\centering
%\includegraphics[width=.85\textwidth]%{figures/z_results_v2.pdf}
%	\caption{Comparison of MORSE relative to several baselines for mGRE $\rightarrow$ MPRAGE translation in Coronal and Sagittal planes. The top row shows the Coronal view, while the bottom row shows the Sagittal one. Note how the traditional 2D diffusion model cDDPM exhibits visible artifacts, while the 2.5D inference of MORSE leads to superior visual, PSNR, and SSIM performance.}
%	\label{fig:z_results}
%\end{figure*}

\begin{table*}[ht]\small
\centering
\setlength{\tabcolsep}{1.5mm}{} 
\renewcommand\arraystretch{1.4}
\begin{tabular}{c|ccclcl}
\hline
Tasks & \multicolumn{6}{c}{mGRE $\rightarrow$ FLAIR}  \\ \hline
planes & \multicolumn{2}{c}{Axial} & \multicolumn{2}{c}{Coronal} & \multicolumn{2}{c}{Sagitta}  \\ \hline
Metrics & PSNR & SSIM & PSNR & SSIM & PSNR & SSIM\\ \hline
U-Net~\cite{oktay2018attention} &21.58  &0.510  &21.80  &0.607  &20.45  &0.677   \\
Pix2Pix~\cite{isola2017image} &24.22  &0.898  &23.09  &0.857  &22.89  &0.843    \\
cDDPM~\cite{ho2020denoising} &\underline{\color[HTML]{00008A}26.75}  &\underline{\color[HTML]{00008A}0.914}  &\underline{\color[HTML]{00008A}24.26}  &\underline{\color[HTML]{00008A}0.872}  &\underline{\color[HTML]{00008A}24.00}  &\underline{\color[HTML]{00008A}0.864}  \\
\hdashline[5pt/5pt]
\textbf{DiffGEPCI} &\textbf{\color[HTML]{D52815}26.92}  &\textbf{\color[HTML]{D52815}0.922}  &\textbf{\color[HTML]{D52815}24.88}  &\textbf{\color[HTML]{D52815}0.889}  &\textbf{\color[HTML]{D52815}24.38}  &\textbf{\color[HTML]{D52815}0.885} \\ \hline
\end{tabular}

\begin{tabular}{c|ccclcl}
\hline
Tasks  & \multicolumn{6}{c}{mGRE $\rightarrow$ MPRAGE} \\ \hline
planes & \multicolumn{2}{c}{Axial} & \multicolumn{2}{c}{Coronal} & \multicolumn{2}{c}{Sagitta} \\ \hline
Metrics  & PSNR & SSIM & PSNR & SSIM & PSNR & SSIM \\ \hline
U-Net~\cite{oktay2018attention}   &25.70  &0.597  &25.45  &0.696  &26.06  &0.784  \\
Pix2Pix~\cite{isola2017image}   &28.31  &0.948  &26.48  &0.912  &26.45  &0.922  \\
cDDPM~\cite{ho2020denoising} &\underline{\color[HTML]{00008A}29.07}  &\underline{\color[HTML]{00008A}0.950}  &\underline{\color[HTML]{00008A}27.57}  &\underline{\color[HTML]{00008A}0.917}  &\underline{\color[HTML]{00008A}27.17}  &\underline{\color[HTML]{00008A}0.925}  \\
\hdashline[5pt/5pt]
\textbf{DiffGEPCI} &\textbf{\color[HTML]{D52815}29.72}  &\textbf{\color[HTML]{D52815}0.956}  &\textbf{\color[HTML]{D52815}28.16}  &\textbf{\color[HTML]{D52815}0.930}  &\textbf{\color[HTML]{D52815}27.95}  &\textbf{\color[HTML]{D52815}0.938}  \\ \hline
\end{tabular}
\caption{Quantitatively comparison of DiffGEPCI and several baselines for mGRE $\rightarrow$ FLAIR and mGRE $\rightarrow$ MPRAGE translation task on testset. The \textbf{\color[HTML]{D52815}best} and \underline{\color[HTML]{00008A} second best} results are highlighted. Note the excellent quantitative performance of DiffGEPCI, which suggests the potential of 2.5D refinement can significantly improve the performance in all three planes. }
\end{table*}

\begin{figure*}[ht]
	\centering
\includegraphics[width=.95\textwidth]{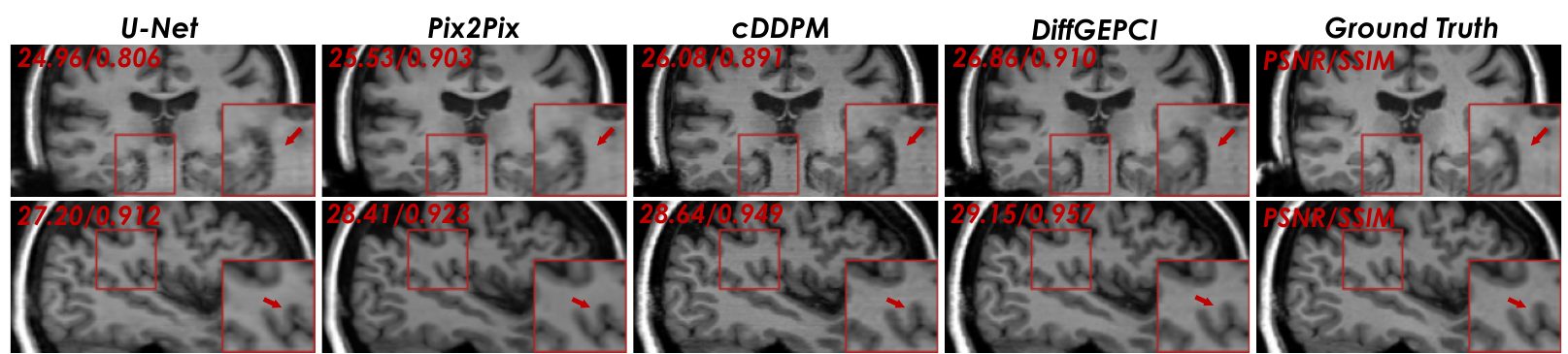}
	\caption{Performance Comparison of various baselines and our proposed DiffGEPCI for mGRE $\rightarrow$ MPRAGE translation in Coronal and Sagittal planes. The top row shows the Coronal view, while the bottom row presents the Sagittal plane.}
	\label{fig:z_results}
\end{figure*}

%\begin{figure}[ht]
%	\centering
%\includegraphics[width=.48\textwidth]%{figures/z_results_part2.pdf}
	%\caption{Comparison of MORSE relative to several baselines for mGRE $\rightarrow$ MPRAGE translation in Sagittal plane.}
	%\label{fig:z_results}
%\end{figure}

\subsection{Comparisons}
We compared DiffGEPCI with three baseline methods. The first method, called U-Net~\cite{oktay2018attention}, corresponds to directly use of a U-Net model to translate the input images into corresponding output images. The second method is pix2pix, which is a cGAN model trained with architecture and loss functions adopted from~\cite{isola2017image}. The third method is cDDPM~\cite{ho2020denoising}, which is a 2D conditional DDPM trained to generate modality outputs based on the condition provided by the other. We used the peak signal-to-noise ratio (PSNR) and structural similarity index (SSIM) as image quality metrics. Before we calculate these quantitative metrics, a brain extraction tool, implemented in ~\cite{bet2}, is used to generate REMs to mask out both skull and background in all modalities.

\subsection{Results}
Figure~\ref{fig:results} shows the visual performance of mGRE image translation to FLAIR and MPRAGE sequences. It is clear that U-Net introduces blurring artifacts and detail loss. Pix2Pix outperform U-Net due to its adversarial nature, but still results in blurry images. The traditional 2D diffusion model excels in recovering details compared with the previous two methods. However, upon closer examination of the generated volume in the  other two planes, cross-slice artifacts become evident, as depicted in  Figure~\ref{fig:z_results}. By using the 2.5D refinement module, DiffGEPCI successfully mitigates these artifacts, attaining state-of-the-art performance across all three planes, as demonstrated in both Figure~\ref{fig:results} and Figure~\ref{fig:z_results}. The quantitative results over the whole
testing set is presented in Table 1, highlighting the quality improvements achieved by DiffGEPCI.

\section{Conclusion}
DiffGEPCI addresses an important problem of generating  translated 3D MRI modalities from a given mGRE signal. Instead of relying solely on a 2D diffusion model that introduces artifacts in orthogonal planes, our novel 2.5D strategy has demonstrated the ability to significantly enhances the image quality of the estimated volumes across all three imaging planes. Through numerical validation using experimentally collected data, DiffGEPCI outperformed several baseline algorithms both quantitatively and qualitatively.

\section{Compliance with Ethical Standards}
This study was performed in line with the principles of the Declaration of Helsinki. Approval was granted by the Ethics Committee of Washington University (Date: 09/05/23/ No: 201906092-1001).

\section*{Acknowledgment}
This work was supported was supported by the NSF CAREER award CCF-2043134 and the the National Institutes of Health (NIH) awards RF1 AG082030 and RF1 AG077658.

% \section*{References}

\bibliographystyle{IEEEbib}
\bibliography{references}

\end{document}